\documentclass[10 pt, letter]{article}
\usepackage{graphicx}
\usepackage{caption}
\usepackage{amsmath}
\usepackage{amsthm}
\usepackage{enumerate}
\usepackage{subfig}
\usepackage{float}
\usepackage{amsfonts}
\DeclareCaptionLabelFormat{Arfken}{\textsc{#1} #2}
\DeclareCaptionLabelSeparator{Arfken}{\quad \ }
\newtheoremstyle{theorems}
  {6pt}
  {18pt}
  {\itshape}
  {}
  {\bf}
  {.-}
  { }
  {\\ \thmname{#1}\thmnumber{ #2}\thmnote{ #3}}

\newtheoremstyle{definitions}
  {6pt}
  {18pt}
  {}
  {}
  {\bf}
  {.-}
  { }
  {\\ \thmname{#1}\thmnumber{ #2}\thmnote{ #3}}

\newtheoremstyle{examples}
  {6pt}
  {18pt}
  {}
  {}
  {}
  {}
  { }
  {\\ \textsc{\thmname{#1}\thmnumber{ #2}:\thmnote{ #3}}}

\newtheoremstyle{example}
  {6pt}
  {18pt}
  {}
  {}
  {}
  {}
  {\newline}
  {\\ \textsc{\thmname{#1}\thmnumber{ #2}:\thmnote{ #3}}\\}

\numberwithin{equation}{section}

\theoremstyle{theorems}

\theoremstyle{definitions}

\theoremstyle{examples}

\theoremstyle{example}

\title{Study of the double non linear quantum resonances in diatomic molecules}
\author{G.V. L\'opez\footnote{gulopez@udgserv.cencar.udg.mx} and J. G. T. Zanudo\footnote{jgtz.fis@gmail.com}\\
\small \emph{Departamento de F\'{i}sica, Universidad de Guadalajara,}\\
\small \emph{Blvd. Marcelino Garc\'{i}a Barragan y Calzada Ol\'{i}mpica, 44200 Guadalajara, Jalisco, Mexico}}
\date{\ \\ \today}
\begin{document}
\maketitle

\begin{abstract}
We study the  quantum dynamics of diatomic molecule driven by a circularly polarized resonant electric field.  We look for a quantum effect due to classical  chaos appearing due to the overlapping of nonlinear resonances associated to the vibrational and rotational motion. We solve the Schr\"odinger equation associated with the wave function expanded in term of proper stationary states, $|n\rangle\otimes|lm\rangle$ (vibrational$\otimes$angular momentum states). Looking for quantum-classic correspondence, we consider the Liouville dynamics in the two dimensional phase space defined by the coherent -like state of vibrational states, and it is found some similarities  when the quantum dynamics is pictured in terms of number and phase operators.
\end{abstract}

\centerline{PACS: 33.40.+f}

\section{Introduction } \label{sec:1}

The study of quantum dynamics in the interval of parameters where classical chaotic behavior occurs is what we call ``Quantum Chaos,  Chaology, or Quantum Manifestation of Chaos" \cite{1}  which deals with some type of quantum manifestation of the classical chaos, mainly associated with the statistical properties of near neighbor levels of energy of the system \cite{1}. In contrast, for quantum systems associated to non chaotic classical ones, it is mostly believed that classical dynamical behavior   must occur for large quantum numbers or high value of the action variable \cite{2} (Rydberg states). In particular, studies of dynamical chaos in atomic and molecular systems has been of great theoretical and practical interest \cite{3}-\cite{12} since not enough integrals of motion are found either in the classical or in its quantum system. Different approaches and studies have been used for the classical \cite{13}-\cite{15} and quantum (quasi-classical region) \cite{16}-\cite{18} cases, and most of them are based on the Morse potential as the inter-atomic interaction \cite{19}.
On the other hand, 
the classical study of the dynamics of atomic and molecular systems has shown that, under certain conditions, these systems are capable of exhibiting a chaotic behavior, even in the case  the system has few degrees of freedom.  In what follows of this introduction and to have a better perspective of the problem, we summarize what Berman et al  \cite{20} did for the classical part of the problem.\\

It is  known that   the dynamics of a diatomic molecule in a resonant circularly polarized electric field can be modeled  
by the Hamiltonian
\begin{equation} \label{eq:1.1}
H(I,\vartheta;p_{\theta},\theta;p_{\varphi},\varphi)= H_0^{(vib)}(I,\vartheta) +H_0^{(rot)}(p_{\theta},\theta; p_{\varphi},\varphi) +H_{int}(I,\vartheta,\theta,t)\\
\end{equation}
where $H_0^{(vib)}$ describes the vibration of the molecule along its axis in terms of the action variable  $I$ and its conjugated angle variable $\vartheta$, $H_0^{(rot)}$ describes the rotation of the molecule around its transversal axis of symmetry in terms of spherical coordinates ($r,\theta,\phi$), and $H_{int}=-\vec d\cdot\vec E(t)$ describes the interaction of the electric dipole moment ($\vec d$) of the molecule with and external electric field ($\vec E(t)$). These term are given explicitly by

\begin{equation} \label{eq:1.2}
  H_0^{(vib)} (I, \vartheta) =\omega_eI - \hbar x_e \omega_e I^2\ ,\quad\quad 
  H_0^{(rot)} (p_{\theta}, \theta;p_{\varphi}, \varphi) = \frac{1}{2\mu r^2}\left(p_{\theta}^2+\frac{p^2_{\varphi}}{\sin^2\theta}\right),
\end{equation}
and
\begin{equation}\label{eq:1.3}
  H_{int}= -E_0d_0-\frac{E_0e_{eff}}{2}\sqrt{\frac{2 I}{\mu \omega_e}}\sin\theta \left[\cos(\vartheta -\varphi + \omega t)+\cos(\vartheta + \varphi - \omega t)\right]. 
\end{equation}
For the derivation of these expressions, the motion of the molecule has been done with respect  the center of mass ($R=(m_1\vec{r_1}+m_2\vec{r_2})/(m_1+m_2)$) and relative ( $\vec{r}=\vec{r_2}-\vec{r_1}$) 
coordinates associated to the diatomic molecule, where $\mu$ is its reduced mass.  The parameter $x_e$ is defined as $x_e=(\hbar/2\omega_e)(d^2H_0^{(vib)}/dI^2)$. The electric field has been chosen as
$ \vec{E(t)}=E_0(\cos \omega t, \sin \omega t, 0)$, 
and the magnitude of the electric dipole moment has been given by
$  d= d_0 + e_{eff}\sqrt{2 I/\mu \omega_e} \cos \vartheta$, 
where $d_0=e_{eff}r_0$, being $e_{eff}$ the effective charge of the molecule and $r_0$ represents the  point of the minimum on the Morse potential [19] which simulates the atom-atom interaction in the diatomic molecule and has been taken up to second order.
The average small vibration oscillation around the equilibrium point $r_o$ is just 
$\int_0^{2\pi} (\Delta r)^2 \ d\vartheta=2I\cos^2\vartheta/\mu\omega_e$, with $\omega_e$ representing the angular frequency of the oscillation of the molecule at first order.  The dynamical system described by this Hamiltonian close to  resonance 
($\omega\approx\omega_e$), and under the condition $\omega\gg |\dot\phi|,|\dot\theta|$  has the following constant of motion
\begin{equation} \label{eq:1.7}
  p_{\varphi}-I=\hbox{constant}=\hbar k,
\end{equation}
The  Hamiltonian  (1.1) can be written in a more suitable form through a change of variable defined by the generatrix function 
 $ F_2(\vartheta,\varphi,\theta; n,k, p; \tau)= \hbar(n+1/2)(\vartheta+\varphi-\omega\tau/\Omega) + \hbar k \varphi + \hbar p\theta $,
which are given by 
\begin{equation}\label{eq:1.8}
\psi  = \vartheta + \varphi - \omega t,\quad\tilde\theta  = \theta, \quad \tilde\varphi   =\ \varphi
\end{equation}
\begin{equation}\label{eq:1.9}
 I  =\hbar (n+1/2),\quad  p_{\theta}   =\hbar p,\quad  p_{\varphi} = \hbar k+I, \quad\tau=\Omega t,
\end{equation}
and the Hamiltonian in this coordinates is written as
\begin{align} \label{eq:1.12}
\widetilde H(n,\psi;p,\theta;k,\varphi)=\ & \hbar (\omega_e-\omega) \left(n+\frac{1}{2}\right) - \hbar x_e \omega_e \left(n+\frac{1}{2}\right)^2 + \beta \left[p^2 + \frac{1}{\sin^2 \theta}\left(k+n+\frac{1}{2}\right)^2\right]\nonumber\\
&-\frac{W}{2}\sqrt{n+\frac{1}{2}}\sin\theta \cos\psi,
\end{align}
where the variable "$n$" is assumed to have continuous values, and  the following definitions have been made: 
$\beta=\hbar^2/2 \mu r_0^2$, and  $W=E_0e_{eff} \sqrt{2\hbar/\mu \omega_e}$.
This Hamiltonian depends only in the conjugate variables $(n,\psi)$ and $(p,\theta)$ since  $\varphi$ is an ignorable variable, and therefore,  $k$ is a constant of motion.  In this way, the Hamilton equations with this variables define the four dimensional classical dynamical system
\begin{align}
\frac{dn}{d\tau}&= - \frac{W}{2}\sqrt{n+\frac{1}{2}}\sin\theta \sin\psi, \nonumber \\
\frac{d\psi}{d\tau}&=\hbar (\omega_e - \omega) - 2\hbar x_e \omega_e \left(n+\frac{1}{2}\right) + \frac{2 \beta}{\sin^2 \theta} \left(k + n + \frac{1}{2}\right) - \frac{W}{4\sqrt{n+1/2}}\sin\theta\cos\psi, \nonumber \\
\frac{dp}{d\tau}&=\frac{2\beta}{\sin^3 \theta}\left(k+n+\frac{1}{2}\right)^2\cos\theta + \frac{W}{2}\sqrt{n+\frac{1}{2}}\cos\theta\cos\psi, \nonumber \\
\frac{d\theta}{d\tau}&=2\beta p.\label{eq:1.14}
\end{align}
This system has its critical points at $p=0$, $\theta=\pi/2$, $\psi=m\pi$ with $m\in{\cal Z}$, and $n=n_i$ with $i=1,2,3$ the roots of a third order polynomial. In the example given by the reference \cite{20},  the parameters associated to the diatomic molecule GeO  \cite{17} are used, 
\begin{align}
  &\hbar \omega_e= 985.8\hbox{ cm}^{-1}, \quad \hbar(\omega_e-\omega)=15\hbox{ cm}^{-1}, \quad \hbar x_e \omega_e = 2.2\hbox{ cm}^{-1}, \quad  \beta=0.48\hbox{ cm}^{-1},\quad\nonumber\\
 & d_0=3.28\hbox{ D}, \quad r_0=1.62 \AA, \quad \mu=13.1\hbox{ amu},  \label{eq:1.15}
\end{align}
where  units have been selected such that $ \hbar \Omega = \frac{3}{2}k_bT= 1\hbox{ cm}^{-1}$ and $ T=0.956\hbox{ K}$, and note that they correspond to a close resonance.  Then, 
there are one center points at $\psi=\pi$ and $n_1\approx 2$ which forms one of the first resonances of the system (for $W=0.05~cm^{-1}$ and $k=0$), and  other resonance which is located near $n_2\approx 1.5$  is due to $\theta$ degree of freedom rather that a critical point of the system. The resonances overlapping Chirikov's criterion \cite{21} for appearing of chaos was verified at $W=0.177~cm^{-1}$, and total chaotic behavior is observed after $W=1.03~cm^{-1}$, see Figure 1.  This result points out that classical chaotic behavior appears within the first two exited states of the associated quantum system. However, the correspondence principle \cite{2} tells us that the quasi-classical behavior of a quantum system is gotten for very large quantum numbers. Therefore, one would not expect any quasi-classical behavior for ground and first exited states of the quantum system.   \\
\begin{figure}[H]
\includegraphics[width=1\textwidth]{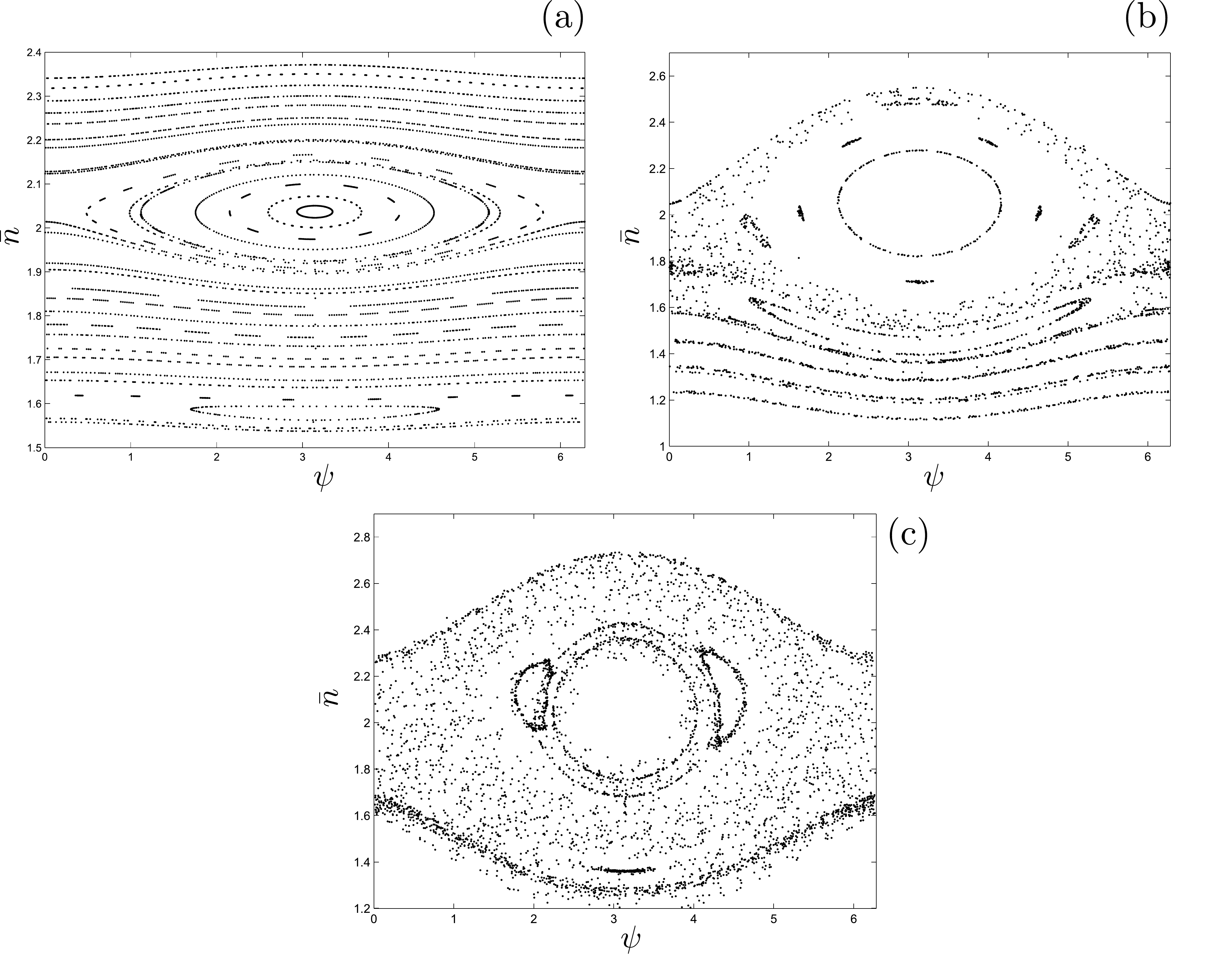}
\centering
  \caption{Poincar\'{e} map for $\theta=\pi/2$ with $\dot \theta > 0$, $k=0$, $\theta_0=1$,and $p_0=0$ and for: (a) $W=0.048\hbox{ cm}^{-1}$; (b) $W=0.68\hbox{ cm}^{-1}$ ; (c) $W=1.03\hbox{ cm}^{-1}$. }        
\label{f:1}
\end{figure}\noindent
In this way,  in this paper we  study the quantum behavior of this system in the region of parameters where this classical chaotic dynamics appears. This behavior  could be important in the study of diatomic molecular clouds for the star born formation or supernova wind shock studies from dying stars \cite{26}, \cite{27}.  We solve the associated Schr\"odinger equation, assuming  the wave function is a linear combination of the stationary states with time depending coefficients and solving numerically the resulting equations for these coefficients , and picture the expectation values of the quantum variables in a phase space-like to look for a similarity with the classical behavior.

\section{Quantum dynamics} \label{sec:2}
\subsection{Quantum Hamiltonian} \label{sec:2.1}
Our goal is to solve the Schr\"odinger equation,
\begin{equation}\label{eq:2.1}
 i\hbar \frac{\partial}{\partial t} |\Psi(t)\rangle = \hat H |\Psi(t)\rangle\ ,
 \end{equation}
 where $ \hat H$ is the Hermitian operator associated to the Hamiltonian,
\begin{align} \label{eq:2.2}
H= \omega_eI - \hbar x_e \omega_e I^2 + \frac{1}{2\mu r_0^2}\left(p_{\theta}^2+\frac{p_{\varphi}^2}{\sin\theta^2}\right)-\frac{E_0e_{eff}}{2}\sqrt{\frac{2 I}{\mu \omega_e}}\sin\theta \cos(\vartheta + \varphi - \omega t)\ .
\end{align}
For this propose,  the following operators are assigned to the observables,
\begin{gather} \label{eq:2.3}
  I \ \longrightarrow \ \hbar\left(aa^{\dagger}+\frac{1}{2}\right),\quad\quad \vartheta \longrightarrow \ \hat \vartheta,\quad\quad L^2=\left(p_{\theta}^2+\frac{p_{\varphi}^2}{\sin^2 \theta}\right) \ \longrightarrow \ \hat{L}^2,\quad\quad \theta \ \longrightarrow \ \hat \theta,\quad\quad \varphi \ \longrightarrow \ \hat \varphi
\end{gather}
where $a$ and $a^{\dagger}=(a)^{\dagger}$ represent the usual ascend and descend operators of the quantum harmonic oscillator. If $\{|n\rangle\}$ and $\{|lm\rangle\}$ are the basis for the harmonic oscillator and the angular moment operators such that
\begin{gather} \label{eq:2.4}
  a|n\rangle=\sqrt{n}|n\rangle, \quad a^{\dagger}|n\rangle=\sqrt{n}|n+1\rangle, \quad a^{\dagger}a|n\rangle=\hat n |n\rangle = n |n\rangle, \quad [a,a^{\dagger}]=1\ ,
\end{gather}
and
\begin{gather} \label{eq:2.5}
\hat L^2|lm\rangle=\hbar^2l(l+1)|lm\rangle,\quad\quad \hat L_z|lm\rangle=\hbar m|lm\rangle,\quad -l\le m\le l
\end{gather}
the action of $\hat \vartheta$ is defined in terms of the phase operator \cite{22}-\cite{24} as
\begin{align} \label{eq:2.6}
  e^{-i\hat \vartheta}:=\frac{1}{\sqrt{aa^{\dagger}}}a, \quad e^{i\hat \vartheta}:=(e^{-i\hat \vartheta})^{\dagger}=a^{\dagger}\frac{1}{\sqrt{aa^{\dagger}}}, \nonumber\\
  \cos \hat \vartheta:=\frac{1}{2}\left(e^{i\hat \vartheta}+e^{-i\hat \vartheta}\right), \quad \sin \hat \vartheta:=\frac{1}{2}\left(e^{i\hat \vartheta}-e^{-i\hat \vartheta}\right).
\end{align}
These operators have the following properties,
\begin{equation} \label{eq:2.7}
  \left[\hat n, \cos \hat \vartheta\right]=i \sin \hat \vartheta, \quad \left[\hat n, \sin \hat \vartheta\right]=-i \cos \hat \vartheta\ , \quad\quad
  e^{i\hat \vartheta}|n\rangle=|n+1\rangle, \quad e^{-i\hat \vartheta}|n\rangle=|n-1\rangle.
\end{equation}
From Eq. (\ref{eq:2.2}) and the definitions (\ref{eq:2.3}) - (\ref{eq:2.5}),  we get the quantum Hamiltonian of the form
$
  \hat{H}= \hat H_0 + \hat W 
 $, 
  where $\hat H_0$ and $\hat W$ are given by
  \begin{equation}\label{eq:2.9}
  \hat H_0= \hbar \omega_e \left(a^{\dagger}a + \frac{1}{2}\right) - \hbar x_e \omega_e \left(a^{\dagger}a + \frac{1}{2}\right)^2+\frac{\beta}{\hbar^2}\hat L^2,
  \end{equation}
  and
  \begin{align}\label{eq:2.10}
  \hat W=& -\frac{W}{2}\sin \hat \theta \left\{\sqrt{\hat n + \frac{1}{2}}\left[\cos \hat \vartheta \cos (\hat \varphi - \omega t) - \sin \hat \vartheta \sin (\hat \varphi - \omega t)\right] \right. \nonumber \\
  &\left.+\left[\cos \hat \vartheta \cos (\hat \varphi - \omega t) - \sin \hat \vartheta \sin (\hat \varphi - \omega t)\right]\sqrt{\hat n + \frac{1}{2}}\right\}.
\end{align}
To construct the Hermitian operator $\hat W$,  we have used the fact that for any operator $\hat A$, the operator $(\hat A+\hat A^{\dagger})/2$ is Hermitian. Using the commutation relations of Eq. (\ref{eq:2.7}), one gets the commutativity of the vibrational and rotational operators , $[f(a,a^{\dagger}),g(\hat \theta,\hat \varphi)]=0$, and we see that $\hbar \hat k = \hat L_z - \hbar (\hat n + 1/2)$ is a constant of motion, that is
\begin{equation} \label{eq:2.11}
  [\hat H, \hbar \hat k]=\left[\hat H, \hat L_z - \hbar \left(\hat n + \frac{1}{2}\right)\right] = 0
  \end{equation}
which is the quantum analogue of Eq. (\ref{eq:1.7}) and has the physical meaning that external electric field's photons excite the rotational and vibrational degree of freedom with the same number of quanta. The main reason for choosing the phase operator as  Eq. (\ref{eq:2.6}) was to be able to get  this quantum constant of motion correctly.
\subsection{Time evolution equations} \label{sec:2.2}
Using the number states $|n\rangle$ and the spherical harmonic states  $|lm\rangle$, we see that $\hat H_0$ is diagonal in the basis $\{|nlm\rangle=|n\rangle\otimes|lm\rangle \}$, and the eigenvalues are given by
\begin{equation} \label{eq:2.12}
  \hat H_0 |nlm\rangle = E_{nl}|nlm\rangle, \quad E_{nl}=\hbar \omega_e \left(n+\frac{1}{2}\right) - \hbar x_e \omega_e \left(n+\frac{1}{2}\right)^2 + \beta l(l+1)\ ,
\end{equation}
where there  is  a $(2l+1)$ degeneration due to quantum number "m". On the other hand, since the relation  $E_{n+1,l}-E_{nl}>0$ must be satisfied, the quantum vibrational number "n" is bounded  \cite{19} in the following way
\begin{align} \label{eq:2.13}
0\le n\le \left[\frac{1}{2 x_e}-\frac{1}{2}\right] ,
\end{align}
where $[x]$ means the integer part of the real number "x".  Therefore, the spectrum is finite. Let us  propose the solution of Eq. (2.1) of the form
\begin{equation} \label{eq:2.14}
  |\Psi(t)\rangle=\sum_{nlm}^{max} D_{nlm}(t)~e^{-itE_{nl}/\hbar} |nlm\rangle\ .
\end{equation}
Now,  substituting  this equation in (\ref{eq:2.1})  and using the orthogonality relation $\langle n'l'm'|nlm\rangle=\delta_{n'n}\delta_{l'l}\delta_{m'm}$, we   get the system of equations for the coefficients as
\begin{equation} \label{eq:2.15}
i\hbar\dot D_{n'l'm'}=\sum_{nlm}^{max}e^{it(E_{n'l'}-E_{nl})/\hbar}D_{nlm}\hat W_{n'l'm';nlm}\ ,
\end{equation}
where the matrix elements $W_{n'l'm',nlm} = \langle n'l'm'|\hat W|nlm\rangle$ are given by
\begin{align}\label{eq:2.16}
  W_{n'l'm',nlm} 
  =& -\frac{W}{2} \left\{\left[\langle n'|\sqrt{\hat n + \frac{1}{2}}\cos \hat \vartheta|n\rangle + \langle n'|\cos \hat \vartheta\sqrt{\hat n + \frac{1}{2}}|n\rangle\right] \langle l'm'| \sin\hat\theta\cos (\hat \varphi - \omega t)|lm\rangle \right. \nonumber\\
  &  \quad \quad \left.- \left[\langle n'|\sqrt{\hat n + \frac{1}{2}}\sin \hat \vartheta|n\rangle + \langle n'|\sin \hat \vartheta\sqrt{\hat n + \frac{1}{2}}|n\rangle \right]\langle l'm'| \sin \hat\theta \sin (\hat \varphi - \omega t)|lm\rangle\right\}\ . \nonumber \\
\end{align}
From the expressions A1 to A7  of the  appendix, this matrix elements can be written as 
\begin{align} \label{eq:2.17}
  W_{n'l'm',nlm}=& \ -\frac{W}{4}\left(\sqrt{n+1/2}+\sqrt{n'+1/2}\right)\left(\delta_{l'}^{l+1}-\delta_{l'}^{l-1}\right)\times \nonumber\\
  &\times\left(e^{-i\omega t}\delta_{n'}^{n+1}\delta_{m'}^{m+1}\frac{c_{l'm'}}{c_{lm}(2l+1)}-e^{i\omega t}\delta_{n'}^{n-1}\delta_{m'}^{m-1}\frac{c_{lm}}{c_{l'm'}(2l'+1)}\right) .
\end{align}
Substituting this expression in Eq. (\ref{eq:2.15}) and using the same dimensionless variables defined in the introduction,  we get the time evolution equation of the coefficients as
\begin{align} \label{eq:2.18}
  i D'_{n'l'm'}= & -\frac{W}{4} \left[ \left(\sqrt{n+1/2}+\sqrt{n-1/2}\right)\frac{c_{lm}e^{i\tau\Omega_{nl,(-),(-)}}}{c_{l-1,m-1}(2l-1)}D_{n-1,l-1,m-1}\right.\nonumber\\
  & \quad \quad \quad -\left(\sqrt{n+1/2}+\sqrt{n-1/2}\right)\frac{c_{lm}e^{i\tau\Omega_{nl,(-),(+)}}}{c_{l+1,m-1}(2l+3)}D_{n-1,l+1,m-1}\nonumber\\
& \quad \quad \quad -\left(\sqrt{n+1/2}+\sqrt{n+3/2}\right)\frac{c_{l-1,m+1}e^{i\tau\Omega_{nl,(+),(+)}}}{c_{lm}(2l+1)}D_{n+1,l-1,m+1}\nonumber\\
& \quad \quad \quad \left.+\left(\sqrt{n+1/2}+\sqrt{n+3/2}\right)\frac{c_{l+1,m+1}e^{i\tau\Omega_{nl,(+),(-)}}}{c_{lm}(2l+1)}D_{n+1,l+1,m+1}\right]\ ,\nonumber\\
\end{align}
where we have made the definitions  $D'=dD/d\tau$,  $ \Omega_{nl,(\pm),(-)}=E_{n\pm1,l-1}-E_{nl}\mp\hbar\omega$, and 
 $ \Omega_{nl,(\pm),(+)}=E_{n\pm1,l+1}-E_{nl}\mp\hbar\omega$.
The selection rules deduced from (\ref{eq:2.18}) are
\begin{equation} \label{eq:2.20}
  \Delta l= \pm 1, \quad  \Delta n = \pm 1\,\quad\hbox{and}\quad \Delta m=\pm 1\ .
\end{equation}
Note that the last two terms of these expressions  are a consequence of the constant of motion (\ref{eq:2.11}).
The time evolution of the coefficients in Eq. (\ref{eq:2.18}) and the selection rules in Eq. (\ref{eq:2.20}) are similar to the  electric dipole transitions in an atom, except with the extra selection rule of $n$. Furthermore, suppose we are initially in a given state $|\Psi(0)\rangle=|n_0 l_0 m_0\rangle$ and we set the frequency $\omega$ to be such that it is almost in resonance with the frequency of an allowed transition, say $|n_f l_f m_f\rangle$ (that is $\hbar\omega\sim E_{n_f,l_f}-E_{n_0,l_0}$). For this case and neglecting the non-resonant transitions,  the equations of motion (\ref{eq:2.18}) becomes
\begin{equation} \label{eq:2.21}
  iD'_0= \alpha e^{i\tau\Omega_r} D_f \quad\hbox{and}\quad  iD'_f= \alpha e^{-i\tau\Omega_r} D_0,
\end{equation}
where  $\alpha$ and $\Omega_r$ are defined as
$\alpha=-\left(\sqrt{n_0+1/2}+\sqrt{n_f+1/2}\right)W c_{l_0,m_0}/4c_{l_f,m_f}(2l_f+1)$ and $
 \Omega_r=E_{n_f,l_f}-E_{n_o,l_o}-\hbar\omega$.
In matrix notation, Eq. (\ref{eq:2.21}) is written as
\begin{equation*}
  i\frac{d}{d\tau}\begin{pmatrix}D_0\\D_f\end{pmatrix}=\alpha \begin{pmatrix}0 &e^{-i\Omega\tau}\\ e^{i\Omega\tau}& 0\end{pmatrix}\begin{pmatrix}D_0\\D_f\end{pmatrix}
\end{equation*}
which in terms of the Pauli operators becomes
\begin{align*}
  i\frac{d}{d\tau}|\psi\rangle&=\alpha\left[\cos\Omega_r\tau \ \hat \sigma_x + \sin \Omega_r\tau \ \hat \sigma_y\right]|\psi\rangle, \quad \hbox{with}\quad |\psi\rangle=\begin{pmatrix}D_0\\D_f\end{pmatrix},
\end{align*}
and this  one is of the form
\begin{align} \label{eq:2.23}
i\frac{d}{d\tau}|\psi\rangle&=\hat H_{at}|\psi\rangle, \quad\hbox{with}\quad \hat H_{at}=\alpha\biggl[\cos\Omega_r\tau \ \hat \sigma_x + \sin \Omega_r\tau \ \hat \sigma_y\biggr],
\end{align}
which is the Schr\"{o}dinger equation for a two level atom introduced in a circularly polarized electromagnetic field \cite{5}.
\subsection{Numerical results} \label{sec:2.3}
\begin{figure}[H]
\includegraphics[width=1\textwidth]{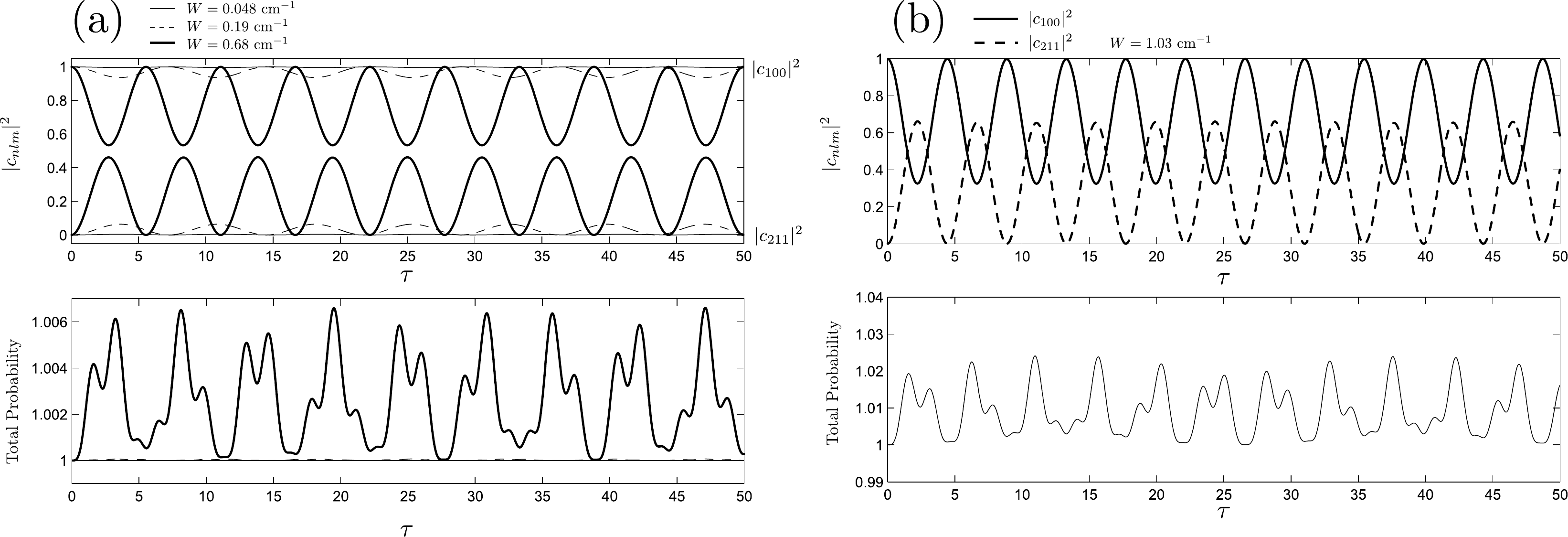}
\centering
    \caption{Time evolution of the total probability, the probability amplitude of the state $|100\rangle$ (upper line) and the state $|211\rangle$ (lower line) for different values of the perturbation: (a) $W=0.048\hbox{ cm}^{-1}$, solid; $W=0.19\hbox{ cm}^{-1}$, dashed; $W=0.68\hbox{ cm}^{-1}$, bold; (b) $W=1.03\hbox{ cm}^{-1}$.}
    \label{fig:A}
\end{figure}

We solve numerically the Eq. (\ref{eq:2.18}), considering only the  coefficients $D_{nlm}$ for $n,l\leq3$.  We use the same parameters used in the classical numerical calculations, Eq. (1.9), which implies to have  a close resonant transition  between the states $|100\rangle$ and $|211\rangle$ with $  |\Omega_{10,(+),(+)}|=0.82$. Higher order of excitation are not
considered since we want to see what happen the states closely related with the classical ones, where classical chaos appears. The results of the numerical simulations are shown in Fig. \ref{fig:A}(a) and Fig. \ref{fig:A}(b) .  The Fig. \ref{fig:A}(a) shows the transition probabilities for small values of $W$ before there is a considerable mixing of states (observe that $|c_{211}|^2<0.5$).  Fig \ref{fig:A}(b) shows the same probabilities but for the value $W=1.03~cm^{-1}$,  and  here one can observe a considerable overlapping of the main transition coefficients involved in the dynamics  (this case  corresponds to have classical chaotic behavior). 
\noindent
Note that  the value of $W$ for classical chaos to appear is  $W_{ch}=0.177\hbox{ cm}^{-1}$, and this one does not coincide with the value for mixing of states ($W\approx0.68\hbox{ cm}^{-1}$), although  they are of the same order of magnitude. We see also that the classical value of the closed classical resonance  suggests overlapping  between quantum states in $n=1$ and $n=2$, as we precisely observe in our simulations, which is consequence of the resonant transition frequency between the states $|100\rangle$ and $|211\rangle$.
\subsection{Quantum phase space pictures} \label{sec:2.4}
\begin{figure}[H]
\includegraphics[width=1\textwidth]{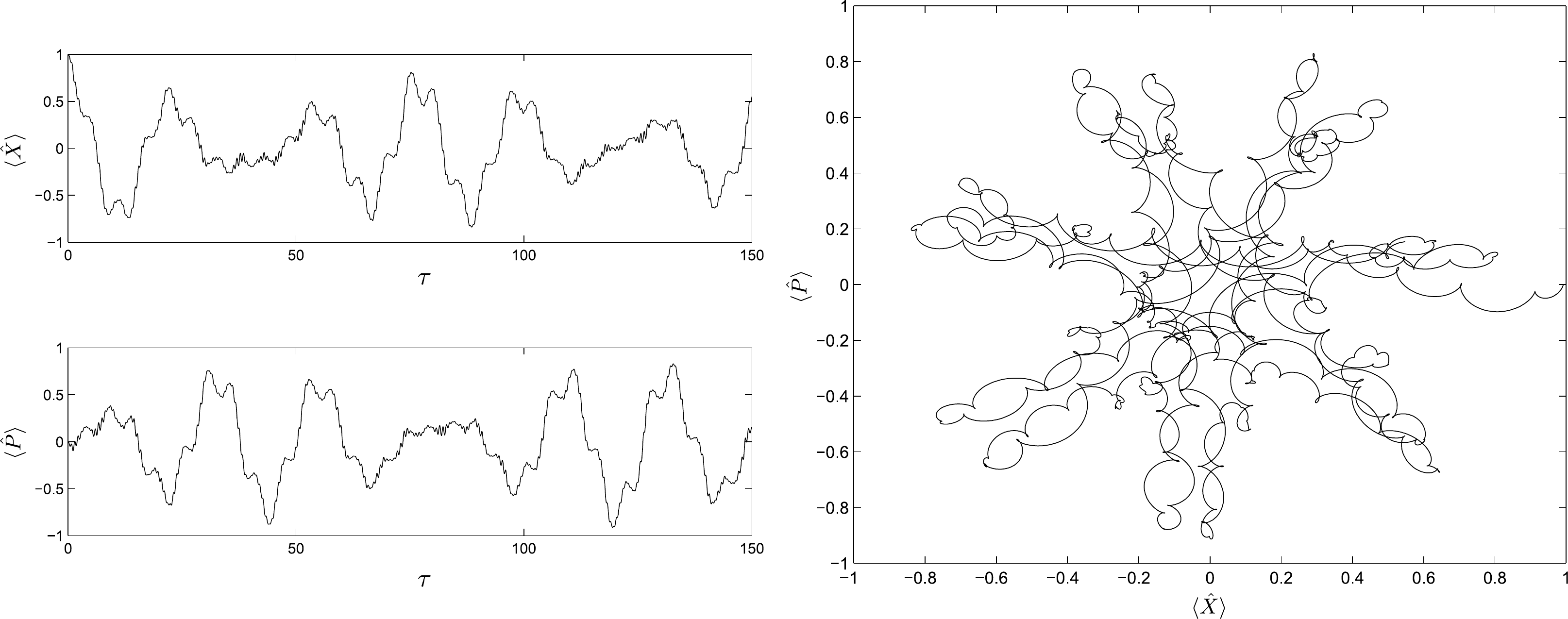}
\centering
    \caption{Phase space like picture with the expectation values variable $\langle X\rangle$ and $\langle P\rangle$ 
    for the initially poission-like distribution, Eq. (\ref{eq:2.25}), using the same parameters as with the classical case. }
\label{fig:B}
\end{figure}

\begin{figure}[H]
\includegraphics[width=1\textwidth]{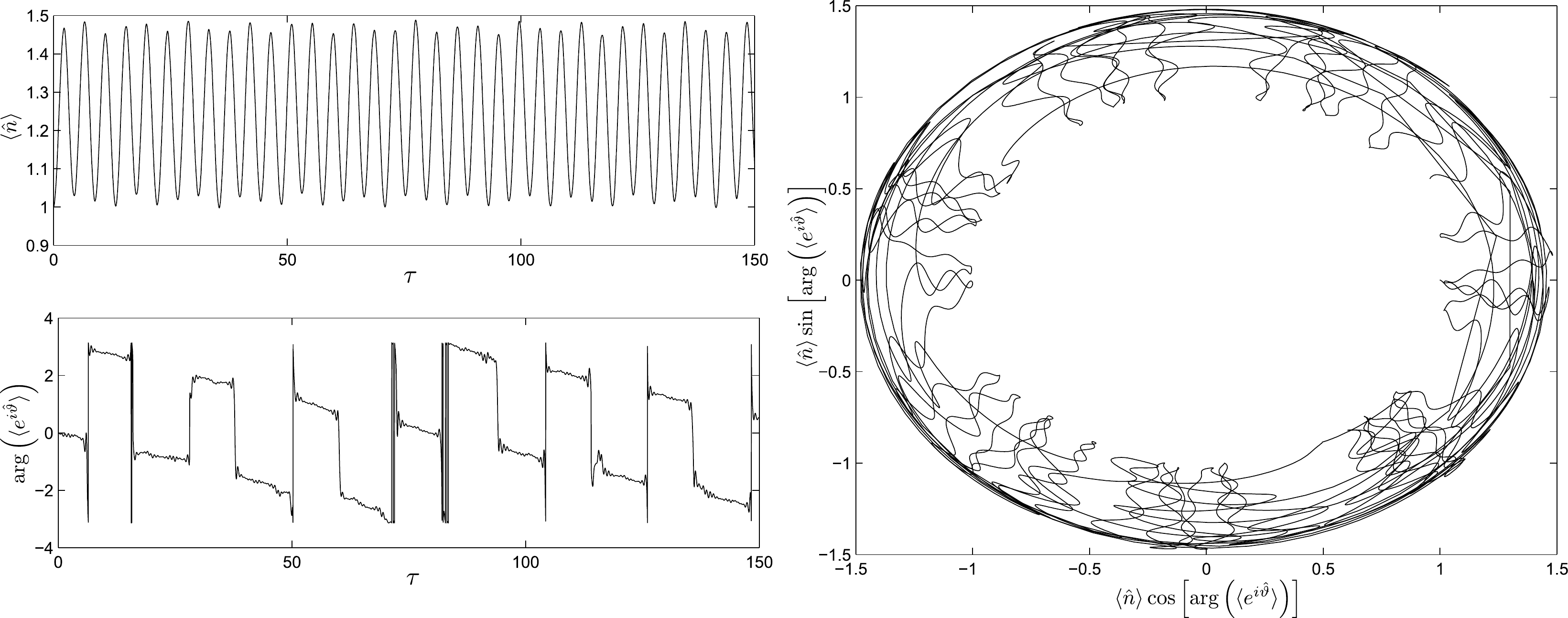}
\centering
    \caption{Phase space like picture  with the same parameters as in the classical case in terms of the expectation values $\langle\hat n\rangle$ and $\langle arg(e^{i\hat \vartheta})\rangle$. }
    \label{fig:C}
\end{figure}

In this section we try two different approaches to get a better  relation between the quantum and classical dynamics. The first and most used approach \cite{22}-\cite{24}, is to use the phase space representation in terms of the expectation value of the dimensionless canonical variables $\hat X$ and $\hat P$
\begin{equation} \label{eq:2.24}
  \hat X= \frac{a + a^{\dagger}}{2},\quad \hat P= \frac{a-a^{\dagger}}{2i}.
\end{equation}
The results of the numerical simulations of this approaches is presented in Fig. \ref{fig:B} , where we used the same parameters of expression (1.15) and $W=1.03~cm^{-1}$.  For the initial state wave function of the system we chose a poisson-like distribution in the coefficients $D_{nl0}$ with the maximum in value in the state $D_{100}$. This initial state is determined by the following coefficients 
\begin{align} \label{eq:2.25}
  D_{000}&=\sqrt{\frac{1.5}{12}},& D_{010}&=\sqrt{\frac{0.2}{12}},& D_{020}&=\sqrt{\frac{0.05}{12}}\nonumber \\
  D_{100}&=\sqrt{\frac{8}{12}},& D_{110}&=\sqrt{\frac{0.4}{12}},& D_{120}&=\sqrt{\frac{0.1}{12}}\nonumber \\
  D_{200}&=\sqrt{\frac{1.5}{12}},& D_{210}&=\sqrt{\frac{0.2}{12}},& D_{220}&=\sqrt{\frac{0.05}{12}}.
\end{align}
 Based in the properties of the coherent states of the harmonic oscillator, this selection not only gives a well defined initial value of the expectation values, but also permits a further study in terms of the Liouville dynamics for both the classical and quantum case \cite{25}. The initially values with quantum number $m=0$ is allowed by   the slow oscillations of $\theta$ and $\varphi$ assumed in the model. Although the dynamics of each variable seem to be stable and similar to each other, the phase space representation does not seem to give any  picture alike to the classical dynamics of the system (see reference [20]), i.e., the phase space  in terms of the canonical variables $\langle\hat X\rangle$ and $\langle\hat P\rangle$ shows no resemblance to the classical dynamics, neither it  seems to suggest any   transition to quasi-classical chaos.\\ \\
\newpage\noindent
In the second approach, we will use  the phase space representation in terms of the expectation value of the phase $\hat \vartheta$ and the number operator $\hat n$. In polar coordinates, the $\langle \hat n \rangle$ will correspond to the radius and  $\hbox{arg}\left(e^{i\hat \vartheta}\right)$ to the angle of the phase space representation of this set of variables,  
$  \hat n= a^{\dagger}a$ and  $e^{i\hat \vartheta}= a^{\dagger}(1/\sqrt{aa^{\dagger}})$.  The expectation values of these variables represents the classical analogous of the variables on Figure 1 above. 
In the upper left plot of Fig. \ref{fig:C} it is shown the time evolution of $\langle n\rangle$ which resemblances to the classical case in terms of the main and different frequencies with which it oscillates. For the  numerical simulations results  presented in these figures  we used the same parameters as in the section \ref{sec:2.3}, and the same initial state (\ref{eq:2.25}).  The phase space picture in terms of the operators $\hat n$ and $e^{i\hat \vartheta}$ seems to have a little bit resemblance with the classical results, perhaps because  the dynamics  of $\hat n$ resembles the classical part. Also, the sudden slow changes of $\hbox{arg}\left(e^{i\hat \vartheta}\right)$ seem to suggest some kind of relation with the resonances of the classical case.
\newpage
\section{Conclusions and comments} \label{sec:5}
We have studied the quantization of a diatomic molecule by solving the Schr\"odinger equation with the known Hamiltonian of the diatomic molecule with a circularly polarized resonant rf-field, written in spherical coordinates (rotations) and angle-action variables (vibrations). The wave function  was expanded in a finite combination  of a proper stationary basis with time dependent coefficients, and the system of equations  for these coefficients was obtained.  Using the same parameters as in the classical case, a near resonant transition between the states $|100\rangle$ and $|211\rangle$ is gotten, which correspond to the closer integer numbers for $n$ where the classical non linear resonances appeared, $n_1 \approx 2$ and $n_2 \approx 1.5$.  The critical value for the mixing between this two states to occur is within the same order of magnitude as the classical value for resonances to overlap.
Using as an initial state a poisson-like distributed wave function in the quantum numbers n and l with maximum in the resonant state $|100 \rangle$, we try two different approaches to see the quantum phase space expectation value dynamics and compare it with the classical case. The usual approach, using the canonical variables $\hat X$ and $\hat P$, fails to provide any intuitive picture of the classical case. On the other hand, the approach using the expectation value of $e^{i \hat \varphi}$ and $\hat n$ suggests  some resemblance and relationship with the classical case. 

Therefore, we have here the following situation, on one hand, the correspondence principle  tells us that we must have the quasi-classical behavior for this quantum system at very large quantum number. However,   due to the quantum dynamics involves  its first few states, quasi-classical chaotic behavior can not be  obtained here, but classical chaotic behavior is obtained just for these first (continues) states.   We do not see how this quasi-classical limit could happen for this quantum system with double nonlinear resonance. So, as one could expect for this case, quantum dynamics does not follow the classical one.\\Ê\\
\section{Acknowledgement}
We want to thank Professor Gennady P. Berman for his comments and suggestions about  this subject, and CONACYT for 
its support with the grand number 0104129.

\newpage
\section{ Appendix}
Matrix elements of some operators
$$
  \langle n'|\sqrt{\hat n + \frac{1}{2}}\cos \hat \vartheta|n\rangle =\frac{1}{2}\sqrt{n'+\frac{1}{2}}\left(\delta_{n',n-1}+\delta_{n',n+1}\right), \eqno(A1)
  $$
 $$
  \langle n'|\cos \hat \vartheta\sqrt{\hat n + \frac{1}{2}}|n\rangle =\frac{1}{2}\sqrt{n+\frac{1}{2}}\left(\delta_{n',n-1}+\delta_{n',n+1}\right), \eqno(A2) 
  $$
 $$
  \langle n'|\sqrt{\hat n + \frac{1}{2}}\sin \hat \vartheta|n\rangle =\frac{1}{2i}\sqrt{n'+\frac{1}{2}}\left(\delta_{n',n+1}-\delta_{n',n-1}\right), \eqno(A3)
  $$
  $$
  \langle n'|\sin \hat \vartheta\sqrt{\hat n + \frac{1}{2}}|n\rangle =\frac{1}{2i}\sqrt{n+\frac{1}{2}}\left(\delta_{n',n+1}-\delta_{n',n-1}\right), \eqno(A4) 
  $$
  $$
  \langle l'm'| \sin\hat\theta\cos \hat \varphi |lm\rangle
  =\frac{1}{2}\left[\frac{\delta_{m'}^{m+1}}{2l+1}\frac{c_{l'm'}}{c_{lm}}\left(\delta_{l'}^{l+1}-\delta_{l'}^{l-1}\right)+\frac{\delta_{m'}^{m-1}}{2l'+1}\frac{c_{lm}}{c_{l'm'}}\left(\delta_{l}^{l'+1}-\delta_{l}^{l'-1}\right)\right], \eqno(A5)
  $$
 $$
  \langle l'm'| \sin\hat\theta\sin \hat \varphi |lm\rangle
  =\frac{1}{2i}\left[\frac{\delta_{m'}^{m+1}}{2l+1}\frac{c_{l'm'}}{c_{lm}}\left(\delta_{l'}^{l+1}-\delta_{l'}^{l-1}\right)-\frac{\delta_{m'}^{m-1}}{2l'+1}\frac{c_{lm}}{c_{l'm'}}\left(\delta_{l}^{l'+1}-\delta_{l}^{l'-1}\right)\right]. \eqno(A6)
$$
The coefficients $c_{lm}$ are given by
$$c_{lm}=\sqrt{\frac{2(l+m)!}{(2l+1)(l-m)!}}\ .\eqno(A7)$$
\newpage
%
%

\end{document}